\documentclass[a4paper]{article}
\usepackage{graphicx}
\newcommand{\be}{\begin{equation}}
\newcommand{\ee}{\end{equation}}
\newcommand{\bd}{\begin{displaymath}}
\newcommand{\ed}{\end{displaymath}}
\newcommand{\bea}{\begin{eqnarray}}
\newcommand{\eea}{\end{eqnarray}}
\newcommand{\bi}{\begin{description}}
\newcommand{\ei}{\end{description}}
\newcommand{\bq}{\begin{quote}}
\newcommand{\eq}{\end{quote}}

\def\S{\Sigma}

\def\ph{\phi}

\begin{document}
\bibliographystyle{unsrt}
\author{Alexander~Unzicker\\
        Pestalozzi-Gymnasium  M\"unchen, Germany \\[0.6ex]
Jan~Preuss\\
        Technische Universit\"at  M\"unchen, Germany \\[0.6ex]
{\small{\bf e-mail:}  aunzicker et web.de}}

\title{A Machian Version of Einstein's Variable Speed of Light Theory}
\maketitle

\begin{abstract}
It is a little known fact that while he was developing his theory of general 
relativity, Einstein's initial idea
was a variable speed of light theory \cite{Einst:07}.
Indeed space-time curvature can be mimicked 
by a speed of light $c(r)$ that depends on the distribution of masses.
 Einstein's 1911 theory was considerably improved by Robert Dicke in
1957, but only recently has the equivalence of the variable speed of light approach to the
conventional formalism been demonstrated \cite{Bro:04}.
Using Green's functions, we show that Einstein's 1911 idea can be expressed in
an analytic form, similar to the Poisson equation. Using heuristic arguments, we
derive a simple formula that directly relates
curvature $w$ to the local speed of light, $w= -c^2 \Delta \frac{1}{c^2}$. In
contrast to the conventional formulation, this allows for a Machian interpretation
of general relativity and the gravitational constant $G$. Gravity, though described
by local equations, has its origin in all other masses in the universe.

\end{abstract}

\section{Introduction: Einstein's unrecognized masterstroke.}

As a direct consequence of his equivalence principle of 1907, Einstein deduced
that light rays both in accelerated and in gravitationally influenced frames must be curved.
His immediate idea was that masses could influence the velocity of light in their vicinity,
much as in optics a medium reduces the local speed of light. He assumed that:
\bq `... the velocity of
light in the gravitational field is a function of the place, we
may easily infer, by means of Huyghens's  principle, that
light-rays propagated across a gravitational field undergo
deflexion.'
\eq
Einstein realized that a variable speed of light must also influence the local units
of time, and in a subsequent paper \cite{Einst:11} in 1911 he stated:
\bq
Nothing forces us to assume [...] the clocks to be running at equal speed.'
\eq
This does not contradict the constancy of the speed of light with respect to Lorentz frames.
With regard to special relativity, he explicitly stated \cite{Einst:12b} that
\bq
`The principle of constancy of the speed of light can be upheld only when one restricts
oneself to regions of constant gravitational potential.'
\eq
The gravitational potential will play an important role in the 
following considerations, being directly related to the speed of light.
However, Einstein's idea remained relatively unknown because of a series of coincidences. 
First and foremost, Eddington's spectacular eclipse observations of 1919 were
seen as a confirmation of the 1915 form of general relativity. 
This overshadowed the fact that Einstein's earlier attempts 
had a merit of their own. Then, because he had not considered length scales, Einstein
derived a wrong value of 0.83 arc seconds with his variable speed of light theory.
Coincidentally, another unsuccessful attempt, the `Entwurf theory' developed in 1913 with Grossmann,
predicted the same wrong value. Thus the variable speed of light attempt is 
easy to confound with that failed idea.

Einstein's theory of 1911 was wrong
in the sense that he did not consider a variation of length scales. This omission
was not obvious prior to the 1919 eclipse results, which seemed to single out the
 geometric formulation of general relativity. It was not until 1957 that
Einstein's error was corrected in a paper \cite{Dic:57} by Robert Dicke, the Princeton astrophysicist
who later became famed for his contribution to the CMB discovery. 
Dicke noted that when he included length scales, the variable speed of light formulation of general
relativity was in agreement with the classical tests, giving an explicit formula for 
the dependence of $c$. Dicke therefore introduced sa variable index of
refraction (\cite{Dic:57}, eq.~5) 
\be 
\frac{c_0}{c}= 1+\frac{2 G M}{r c^2} =: 1+2 \alpha  \label{dic} 
\ee 
whereby $c_0$ is the velocity far from all masses. Given that the small right term 
depends on the Sun, Dicke discovered a relation with Mach's principle: He suspected that the 
main part of the right term, 1, might have 'its origin in the remainder 
of the matter in the universe'. The
intriguing consequences for cosmology with respect to Dirac's 
large number hypothesis, and Dicke's further considerations are
addressed in \cite{Unz:07c}. Although Dicke still had difficulties in reproducing the
same results as he obtained for general relativity, the equivalence of both formulations was 
proven by \cite{Bro:04}. It is worth mentioning that the notion of a variable 
speed of light is present in the conventional formalism of general relativity as
well, usually denoted as `coordinate velocity.' (For numerous references, 
see \cite{Bro:04},  ref.[70] herein.)

Since the considerable differences in the mathematical formalisms have obviously hindered
the general acceptance of the variable speed of light formulation,
we show in the following that Dicke's idea can be constituted in a form much closer to
Einstein's classical equations. In 1912, Einstein \cite{Einst:12} (eq. 5b) also suggested 
an analytic form resembling the Poisson equation, 
but in a quite different manner.

\section{General relativity and variable speed of light}
Before we modify Dicke's idea, it is worth mentioning that other researchers have
tried to incorporate Mach's principle into a theory of gravitation. A couple of years earlier than Dicke,
Sciama \cite{Sci:53} derived a hypothesis on the gravitational constant,
\be
G= \frac{c^2}{\sum_i \frac{m_i}{r_i}}. \label{Sciama53}
\ee
The coincidence (\ref{Sciama53}) was, however, taken as an approximation.
How Sciama motivated his approach based on Mach's 
principle is discussed in \cite{Unz:07c}.
 As early as 1925, Erwin Schr\"odinger made similar 
suggestions in a far-sighted paper \cite{Schr:25}.

\subsection{Recovery of Newton's law}

We first show that that Dicke's and Sciama's idea are both embedded in the formula
\be
c^2= \frac{c_{0}^2}{\sum_i \frac{m_i}{r_i}}, \label{gpot2}
\ee
$c_0$ being a constant velocity at an infinite distance. Since in (\ref{dic}),
the right term is small, $c_0^2/c^2 \approx 1+4 \alpha$ holds. If one assumes the `1' arising
from a corresponding sum,  Dicke's suggestion is equivalent
to a Newtonian gravitational potential in the form 
\be 
\ph_{Newton} = \frac{1}{4}c^2.
\ee 
Newton's law for the acceleration of a test mass at $r=0$ can be recovered by 
\be \vec a (\vec r)
= - \frac{1}{4} \nabla c(\vec r)^2 = \frac{c_0^2}{4 \S^2}
\sum_i \frac{\vec r_i}{r_i^3},
\ee
where $\S$ is an abbreviation for $\sum_i \frac{m_i}{r_i}$. When substituting $c_0^2$,
\be
a= \frac{c^2}{4 \S} \sum_i \frac{\vec e_i}{r_i^2} \label{newton}
\ee
follows, yielding the inverse-square law, while the gravitational constant is expressed by
$G= \frac{c^2}{4 \S}$.
Note that $c_0$ does not appear any more in (\ref{newton}), thus this is a form
of Dicke's idea where the Newtonian force is 
perceived in local, dynamic units.

\subsection{The Einstein-Dicke variable speed of light theory in analytic form}

We now present the Sciama-Dicke idea of a Machian law of gravity in a new analytic form.
Making use of Einstein's gravitational constant
\be
\kappa ~=~ \frac{8\pi\,G}{c^4},
\ee
Sciama's (modified) hypothesis
\be
G= \frac{c^4 \kappa}{8 \pi}= \frac{c^2}{4 \sum \frac{m_i}{r_i}},
\ee
can be rewritten as
\be
\frac{1}{c^2} = \frac{\kappa}{2 \pi} \sum \frac{m_i}{r_i}.
\ee
The corresponding integral form is
\be
\frac{1}{c^2} = \frac{\kappa}{2 \pi} \int d \vec r' \frac{\rho}{|\vec r' -\vec r|}. 
\ee
Taking the spatial Laplace operator on both sides yields
\be
\nabla_r^2 \frac{1}{c^2} = \frac{\kappa}{2 \pi} \int d \vec r' \nabla_r^2 \frac{\rho}{|\vec r' -\vec r|}.
\label{intform}
\ee
Using the fact that
\be
- \frac{1}{4 \pi} \Delta \frac{\rho}{|\vec r' -\vec r|} ~=~\rho \delta^3 (\vec r' -\vec r),
\ee
Equation (\ref{intform}) transforms to
\be
\Delta \frac{1}{c^2} ~=~ - 2 \kappa \rho. \label{kapparho}
\ee
It is interesting to compare this with Einstein's 1912 proposal \cite{Einst:12b}, $\Delta c = \kappa c \rho$,
considered `the simplest equation of this kind' by Einstein. Evidently, he had seen the need
to express his idea in a local form, though his solution is obviously not the only one.
However, such a local formulation of the Sciama-Dicke hypothesis suggests further modifications that
shed light on the Machian nature of the approach taken by Dicke.

\subsection{The elimination of Newton's constant $G$.}

Since neither Dicke's nor Sciama's approach has beenbuilt into a complete theory,
(\ref{kapparho}) should not be considered as a rigorous consequence, but rather as a 
starting-point for what was behind Dicke's and Sciama's efforts, namely a Machian approach to
gravity.

Einstein's gravitational constant $\kappa=
\frac{8 \pi G}{c^4}$ possesses an interesting property.
 When we look at its physical units $\frac{s^2}{m kg}$.
There have been several theoretical\footnote{Not to be confused with
the experimental efforts to redefine the kilogram.} approaches to redefine the unit of mass $kg$,
which is, from a fundamental perspective, an unjustified quantity. Julian Barbour \cite{Bar}, for example,
suggested that the unit of mass could as well be expressed as an inverse acceleration $\frac{s^2}{m}$, 
since it is the nature of an (inertial) mass to resist acceleration. If we do so,
we imply the validity of Newton's third law by definition. $\kappa$ therefore can be seen 
 as a conversion factor that transforms the regular unit $kg$ into a Machian unit $\frac{s^2}{m}$. 
This has further implications, such as the unit of energy becoming $m$ instead
of $kg \frac{m^2}{s^2}$. 

Considering the preliminary form of (\ref{kapparho}), one may replace the mass density $\rho$ 
by an energy density $w$. Then, multiplying (\ref{kapparho}) by $c^2$, allows the form
\be
-c^2 \Delta \frac{1}{c^2} ~=~ w, \label{1overc2}
\ee
$w$ being the energy density with units $\frac{1}{m^2}$, obviously related to the unit
of the Riemannian curvature tensor. 
If we remember Dicke's original idea, this yet  more simple form is already
an implementation of Mach's principle. In (\ref{1overc2}), the gravitational constant $G$
has disappeared completely, as suggested by Dicke. So far we have considered the static case
only. In general relativity, as in all other circumstances, the time derivative that corresponds to
the Laplacian (and its extension, the Riemann tensor) must have an extra factor $\frac{1}{c^2}$. 
Thus without formal justification, one may assume that the relativistic generalization of (\ref{1overc2})
must have the form
\be
\frac{\partial^2}{\partial t^2} \frac{1}{c^2} -c^2 \Delta \frac{1}{c^2} ~=~ w. \label{cwave}
\ee
Despite being a nonlinear equation, the linear approximation of a locally constant $c$
is equivalent to a wave equation in empty space. A number of questions remain 
for further investigation. It will be interesting to see how (\ref{cwave})
matches the gravitational wave predictions. Another issue is the role of the metric. 
In the conventional formalism, the speed of light $c$ enters the metric merely as a conversion
factor, whereas here it seems that the metric itself is a function of $c$. If it can be 
shown that (\ref{cwave}) is an equivalent description of the phenomenology of general relativity,
it would be a strikingly simple form of Einstein's theory.

\section{Outlook}
These purely heuristic considerations can only be seen as a first step toward incorporating
Mach's principle into a theory of gravity. Yet they constitute a novel way to link Dicke's
alternative theory of gravity of 1957 (not to be confused with the Brans-Dicke theory)
with Einstein's variable speed of light attempt in 1911. The content of
(\ref{1overc2}) is certainly an unusual form of 
Ernst Mach's idea of gravity having its origin in all other masses in the universe.
The idea had always fascinated Einstein, though he never succeeded in incorporating it into 
general relativity. We have shown that his own 1911 idea,
widely unknown in contemporary research, potentially offers an intriguing
explanation of gravity.

The big epistomological difference between conventional and Machian
gravity is that the latter allows for the elimination of a fundamental constant,
Newton's constant $G$. All revolutionary developments in physics have reduced the number
of fundamental constants while expressing one constant in terms of others. Calculating 
$G$ by the mass distribution of the universe, as first suggested by Dicke, is therefore 
a radical idea that needs to be evaluated with care.

\paragraph{Acknowledgement.}
The authors thank for various inspiring discussions with Karl Fabian.


\end{document}